\documentclass[aps,prl,twocolumn,superscriptaddress]{revtex4-1}

\usepackage{fancyhdr} 
\usepackage{color}
\usepackage{hyperref} 

\usepackage{verbatim} 
\usepackage{amsmath} 
\usepackage{amsthm} 
\usepackage{amssymb}	
\usepackage{graphicx} 

\newcommand{\ket}[1]{\big| #1 \big\rangle} 
\newcommand{\bra}[1]{\big\langle #1 \big|} 
\newcommand{\Z}{\mathbb{Z}} 
\newcommand{\disavg}[1]{\overline{#1}} 
\newcommand{\expec}[1]{\left< #1 \right>} 
\newcommand{\diag}{\mathrm{diag}} 
\newcommand{\abs}[1]{\left| #1 \right|} 
\newcommand{\proj}[1]{\ket{#1}\bra{#1}} 
\newcommand{\Tr}{\mathrm{Tr}} 

\usepackage{pdfpages} 
\makeatletter
\AtBeginDocument{\let\LS@rot\@undefined}
\makeatother

\begin{document}

\title{Floquet Quantum Criticality}

\author{William Berdanier}
\email[]{wberdanier@berkeley.edu}
\affiliation{Department of Physics, University of California, Berkeley, CA 94720, USA}

\author{Michael Kolodrubetz}
\affiliation{Department of Physics, University of California, Berkeley, CA 94720, USA}
\affiliation{Materials Sciences Division, Lawrence Berkeley National Laboratory, Berkeley, CA 94720, USA}
\affiliation{Department of Physics, The University of Texas at Dallas, Richardson, Texas 75080, USA}

\author{S. A. Parameswaran}
\altaffiliation{On leave from: Department of Physics and Astronomy, University of California Irvine, Irvine, CA 92697, USA.}
\affiliation{The Rudolf Peierls Centre for Theoretical Physics, University of Oxford, Oxford OX1 3PU, UK}

\author{Romain Vasseur}
\affiliation{Department of Physics, University of Massachusetts, Amherst, Massachusetts 01003, USA}

\date{\today}

\begin{abstract}
We study transitions between distinct phases of one-dimensional periodically driven (Floquet) systems. We argue that these are generically controlled by infinite-randomness fixed points of a strong-disorder renormalization group procedure. Working in the fermionic representation of the prototypical Floquet Ising chain, we leverage infinite randomness physics to provide a simple description of Floquet (multi)criticality in terms of a new type of domain wall associated with time-translational symmetry-breaking and the formation of `Floquet time crystals'. We validate our analysis via numerical simulations of free-fermion models sufficient to capture the critical physics.
\end{abstract}
\maketitle

The assignment of robust phase structure to periodically driven quantum many-body systems is among the most striking results in the study of non-equilibrium dynamics~\cite{khemani_prl_2016}. There has been dramatic  progress in understanding such `Floquet' systems, ranging from proposals to  engineer new states of matter via the drive~\cite{RevModPhys.89.011004,doi:10.1080/00018732.2015.1055918,PhysRevLett.116.205301,0953-4075-49-1-013001,PhysRevX.4.031027,PhysRevB.82.235114,Lindner:2011aa,PhysRevX.3.031005,1367-2630-17-12-125014,PhysRevB.79.081406,Gorg:2018aa} to the classification of driven analogs of symmetry-protected topological  phases (`Floquet SPTs')~\cite{PhysRevB.92.125107,PhysRevB.93.245145,PhysRevB.93.245146,PhysRevB.93.201103,potter_classification_2016,PhysRevB.94.125105,PhysRevB.94.214203,PhysRevB.95.195128}. These typically require that the  system under investigation possess one or more microscopic global symmetries. In addition, {\it all} Floquet systems share an invariance under time translations by an integer multiple of their drive period. Unlike the continuous time translational symmetry characteristic of undriven Hamiltonian systems~\cite{PhysRevLett.109.160401,PhysRevLett.111.070402,PhysRevLett.114.251603}, this discrete symmetry may be spontaneously broken, 
leading to a distinctive dynamical response at rational fractions of the drive period --- a phenomenon dubbed `
time crystallinity'~\cite{pi-spin-glass,else_floquet_2016,PhysRevX.7.011026,PhysRevLett.118.030401,PhysRevB.96.115127,PhysRevLett.119.010602}. The time translation symmetry breaking (TTSB) exhibited by Floquet time crystals is stable against perturbations that preserve the periodicity of the drive, permitting generalizations of notions such as broken symmetry and phase rigidity to the temporal setting. Experiments have begun to probe these predictions in well-isolated systems such as ultracold gases, ion traps~\cite{Zhang:2017aa}, nitrogen-vacancy centers in diamond~\cite{Choi:2017aa}, and even spatially ordered crystals~\cite{1802.00457,1802.00126}. 

In light of these developments, 
it is desirable to construct a theory of Floquet (multi-)critical points between distinct Floquet phases. Ideally, this should emerge as the fixed point of a coarse-graining/renormalization group procedure, enable us to identify critical  degrees of freedom, especially those responsible for TTSB, and allow us to compute the critical scaling behavior.

Here, we  develop such a theory for a prototypical Floquet system: the  driven random quantum Ising chain.  Extensive analysis has shown that this model hosts four phases~\cite{khemani_prl_2016,pi-spin-glass}. Two of these, the paramagnet (PM) and the spin glass (SG), are present already in the static problem~\cite{PhysRevB.88.014206,PhysRevLett.112.217204,PhysRevX.4.011052}. A third, the $\pi$ spin glass/
time crystal,  has spatiotemporal long-range order and subharmonic bulk response at half-integer multiples of the drive frequency. This phase, and its Ising dual --- the $0\pi$ paramagnet, which also exhibits TTSB but only at the boundaries of a finite sample --- are unique to the driven setting. A precise understanding of the (multi)critical points  between these distinct Floquet phases accessed by tuning drive parameters is the subject of this work.

Our approach relies on the presence of quenched disorder, required for a generic {periodically-driven} system to have Floquet phase structure rather than thermalize to a featureless infinite-temperature state~\cite{PONTE2015196,PhysRevLett.114.140401,PhysRevLett.115.030402,ABANIN20161}. 
We argue that transitions between distinct one-dimensional Floquet phases are then best described in terms of an infinite-randomness fixed point accessed via a strong-disorder real space renormalization group procedure. In the non-equilibrium setting, the stability of infinite-randomness fixed points against thermalization via long-range resonances remains a topic of debate~\cite{PhysRevB.95.155129,PhysRevLett.119.150602,Ponte20160428}. However, even if unstable, we expect that they will control the dynamics of prethermalization relevant to all reasonably accessible experimental timescales~\cite{PhysRevLett.115.256803,KUWAHARA201696}.

The universality of our analysis turns on the fact that, in the vicinity of such infinite-randomness critical points, a typical configuration of the system can be viewed as being composed of domains deep in one of two proximate phases~\cite{PhysRevLett.69.534,PhysRevB.50.3799,PhysRevB.51.6411,PhysRevLett.89.277203,PhysRevB.61.1160,PhysRevX.5.031032}. Transitions that do {\it not}  involve TTSB  (i.e., the SG/PM or $0\pi$PM/$\pi$SG transitions) map 
to the static (random) Ising critical point and can be understood in similar terms. 
In contrast, transitions that involve the onset of TTSB in the bulk (PM to $\pi$SG) or at the boundary (SG to $0\pi$PM) can be understood in terms of a new class of domain wall special to driven systems, that separate regions  driven at a frequency primarily near  0 or near $\pi$ --- a picture we verify numerically. When the Ising model is rewritten as a fermion problem, this picture yields a simple description of Floquet criticality in terms of domain walls that bind Majorana states at quasienergy $0$ or $\pi$, allowing us to further study the multicritical point where all four phases meet.

\paragraph{Model.} 

\begin{figure}
\includegraphics[width = \columnwidth]{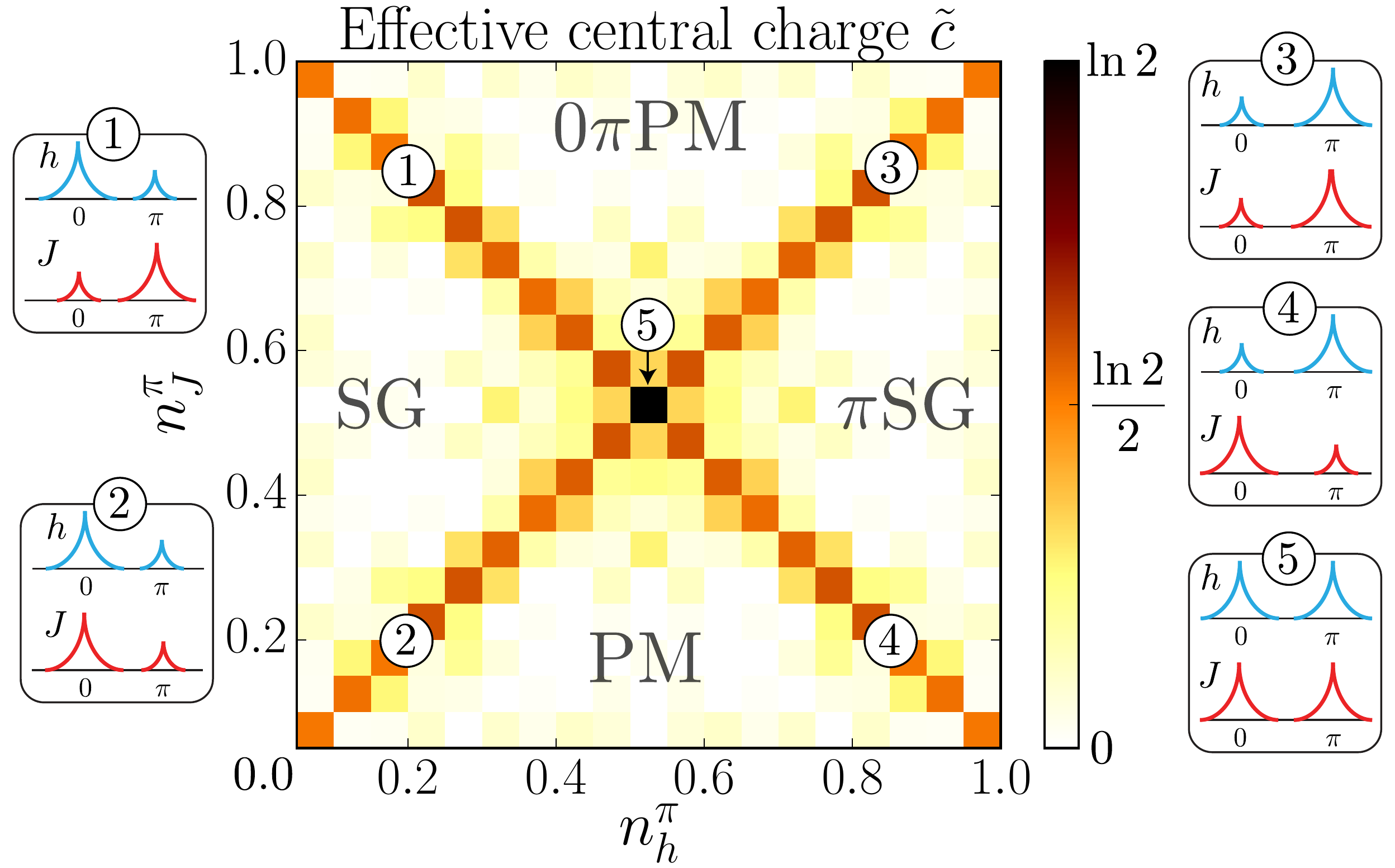}
\caption{\label{fig:phase_diagram} {Phase diagram deduced by fitting ``effective central charge'' from entanglement scaling (see Fig. 3 for details). Insets: sketches of infinite-randomness coupling distributions along the critical lines (1-4) and at the multicritical point (5).}} 
\end{figure}

Floquet systems are defined by a time-periodic Hamiltonian $H(t) = H(t+T)$.  For reasons similar to Bloch's theorem, eigenstates satisfy  
$\ket{\psi_\alpha (t)} = e^{-i E_\alpha t} \ket{\phi_\alpha(t)}$, where $\ket{\phi_\alpha(t+T)} = \ket{\phi_\alpha(t)}$ and we set $\hbar =1$~\cite{PhysRev.138.B979,PhysRevA.7.2203}. In contrast to the case of static Hamiltonians, the quasi-energies $E_\alpha$ are only defined modulo $2\pi/T$, voiding the notion of a `ground state'.

 An object of fundamental interest is the single-period evolution operator or Floquet operator 
$F \equiv U(T)$.
If disorder is strong enough,  $F$ can have an extensive set of local conserved quantities. 
 This implies area-law scaling of entanglement in Floquet eigenstates, and consequently the absence of thermalization~\cite{doi:10.1146/annurev-conmatphys-031214-014726}.

Unlike  generic (thermalizing) Floquet systems, such many-body localized (MBL) Floquet systems retain a notion of phase structure to infinitely long times. For concreteness, we focus on the driven quantum Ising chain, the simplest Floquet system that hosts uniquely dynamical phases. The corresponding Floquet operator is
\begin{equation}
F = {\rm e}^{-i \frac{T}{2} \sum_i J_i  \sigma_i^z \sigma_{i+1}^z + U \sigma^z_i \sigma^z_{i+2} } {\rm e}^{-i \frac{T}{2} \sum_i h_i \sigma_i^x + U \sigma^x_i \sigma^x_{i+1} },\!\!\! 
\label{eqModel}
\end{equation} 
where $\sigma^\alpha_i$ are Pauli operators. Here $J_i$ and $h_i$ are uncorrelated  random variables, 
and $U$ corresponds to small interaction terms that respect the ${\mathbb Z}_2$ symmetry of the model generated by $G_{\text{Ising}} = \prod_i \sigma_i^x$. {For specificity, 
we  draw couplings $h,J$ randomly with probability $n_\pi^{h,J}$ from a box distribution of maximal width about $\pi$, namely $[\pi/2,3\pi/2]$, and with probability $n_0^{h,J} = 1-n_\pi^{h,J}$ from a box distribution of maximal width about $0$, namely $[-\pi/2,\pi/2]$. The reasons for this parametrization will become evident below.}
$F$ corresponds to an interacting transverse-field Ising model where for $U=0$  we stroboscopically alternate between field and bond terms. It is helpful to perform a Jordan-Wigner transformation to map bond and field terms to Majorana fermion hopping terms, yielding a $p$-wave free fermion superconductor with density-density interactions given by $U$. In the high-frequency limit $T \to 0$, we can rewrite $F =  e^{-i H_FT}$ by expanding and re-exponentiating order-by order in $T$
and the Floquet Hamiltonian $H_F$ recovers a static Ising model.  We work far from this limit, setting $T=1$. 

\begin{figure}[t!]
\includegraphics[width =  1. \columnwidth]{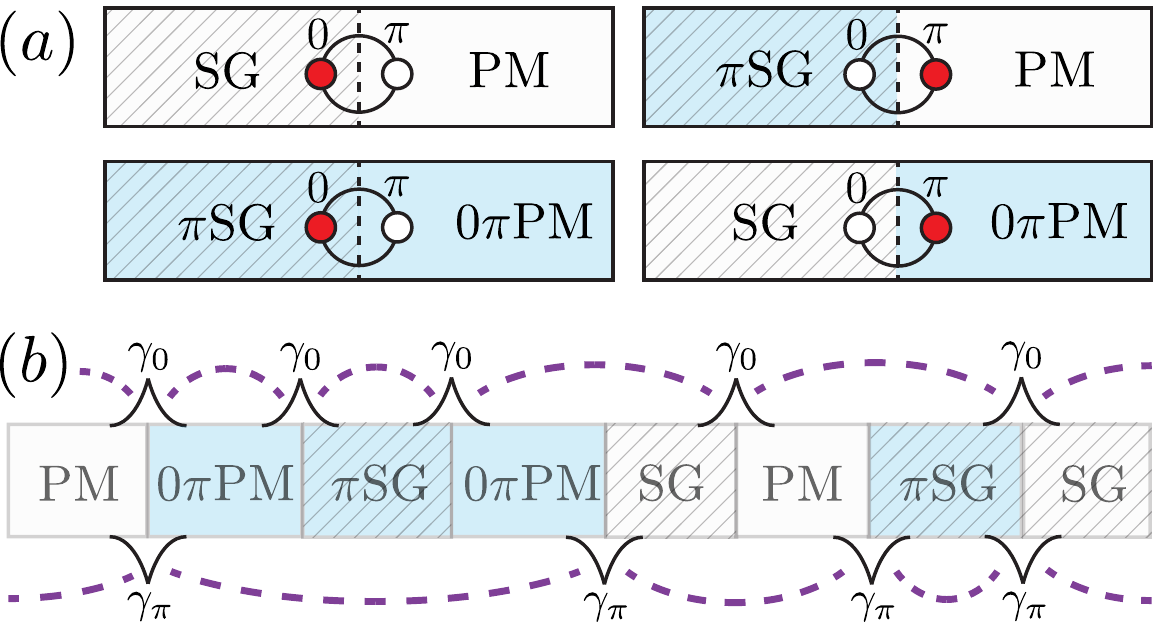}
\caption{\label{fig:cartoons} (a) Domain walls (DWs) between proximate phases of the driven Ising model.  In fermionic language, these host topological edge states at either 0 or $\pi$ quasienergy (red). Blue regions exhibit bulk/boundary  time-translational symmetry breaking (TTSB), and hatched regions have bulk spin glass order. (b) A typical multicritical configuration. Tunneling between DW states $\gamma_{0,\pi}$ yields two independent chains around $0$ and $\pi$ quasienergy.
}
\end{figure}

\paragraph{{Phases and Duality.}} 
{Observe that $(n_\pi^h, n_\pi^J) =\frac{1}{\pi}({\overline{h_i}},{\overline{J_i}})$, where the bars denote disorder averages, and hence tune between phases of model \eqref{eqModel} analogously to $h, J$ in the clean case.} 
The four phases are {summarized in the phase diagram in Figure~\ref{fig:phase_diagram}}. 
The trivial Floquet paramagnet (PM) breaks no symmetries and has short range spin-spin correlations. The spin glass (SG) spontaneously breaks Ising symmetry with long-range spin correlations in time, or equivalently localized edge modes at 0 quasienergy in the fermion language. These two phases are connected to the undriven paramagnet and ferromagnet/spin glass phases of the random Ising model~\cite{PhysRevB.88.014206,PhysRevLett.112.217204,PhysRevX.4.011052}. Unique to the Floquet system are the $\pi$-spin glass ($\pi$SG) and the $0\pi$ paramagnet ($0\pi$PM). The $\pi$SG  spontaneously breaks both Ising and time translation symmetry in the bulk. Often referred to as a ``
time crystal''~\cite{khemani_prl_2016,else_floquet_2016,PhysRevLett.118.030401}, it maps to a fermion phase with localized Majorana edge modes at $\pi$ quasienergy~\cite{PhysRevLett.106.220402}.
Finally, the $0\pi$PM has short range bulk correlations but also boundary TTSB; its fermion dual has both $0$ and $\pi$ Majorana edge modes and is a simple example of a Floquet SPT. In the fermion language, domain walls between these different phases host either $0$ or $\pi$ Majorana bound states (Fig.~\ref{fig:cartoons}a)  central to the infinite-randomness criticality discussed below.  

The absence of energy conservation in the Floquet setting admits two new  {eigenstate-preserving} changes of parameter to \eqref{eqModel}. The transformations $J_j \mapsto J_j + \pi$ and $h_j \mapsto h_j + \pi$ both separately map $F$ onto another interacting Ising-like Floquet operator with precisely the same eigenstates~\cite{supmat}, but possibly distinct quasienergies: $J_j \mapsto J_j+\pi$ preserves $F$ exactly (up to boundary terms), while $h_j \mapsto h_j + \pi$ sends $F \mapsto FG_{\rm Ising}=G_{\rm Ising}F$. {Note that, despite not changing bulk properties of the eigenstates, these transformations map the PM to the $0\pi$PM and the SG to the $\pi$SG respectively.} Additionally,  a global rotation about the $y$ axis takes $h_j\mapsto -h_j$. Below, we fix phase transition lines by combining
 these {\it Floquet symmetries} with the usual Ising bond-field duality that exchanges $h$ and $J$ (and hence SG and PM in the static random case).
  
\paragraph{Infinite-randomness structure.} In analogy with the critical point between PM and SG phases in the static random Ising model (both at zero temperature and in highly excited states), we expect that the dynamical Floquet transitions of~\eqref{eqModel} are controlled by an infinite-randomness fixed point  (IRFP) of a real space renormalization group (RSRG) procedure. At a static IRFP,  the distribution of the effective couplings broadens without bound under renormalization, so the effective disorder strength diverges with the RG scale. A typical configuration of the system in the vicinity of such a transition can be viewed as being composed of puddles deep in one of the two proximate phases, in contrast with  continuous phase transitions in clean systems~\cite{PhysRevB.61.1160,PhysRevX.5.031032}.

 {In order to generalize this picture to the Floquet Ising setting we must identify appropriate scaling variables. For $J_i, h_i \ll \pi$ we recover the criticality of the static model controlled by an IRFP  if $J_i$ and $h_i$ are drawn from the same distribution.  In this case, the relevant operator at the critical point controls the asymmetry between the $J_i$, $h_i$ distributions. At static IRFPs, critical couplings are power-law distributed near $0$.  The absence of energy conservation in the Floquet setting  complicates this picture since there is no longer a clear notion of `low' energies. However, a natural resolution is to allow for fixed-point couplings to be symmetrically and power-law distributed around {\it both} $0, \pi$ quasienergy (or more generally, all quasienergies that can be mapped to $0$ by applying Floquet symmetries of the drive). This introduces a new parameter for Floquet-Ising IRFPs, namely the fractions $n_0$ and $n_\pi$ of couplings near $0$ and $\pi$, respectively. Evidently, we have $n_0=1 - n_\pi$. We will show  
 that there is a new type of IRFP specific to the Floquet setting for $n_\pi = 1/2$, where the criticality is tuned by the asymmetry between the distributions at $0$ and $\pi$ quasienergy, at {\it fixed} values of the  $J_i$ -
 $h_i$ distribution asymmetry.
  
\paragraph{Emergent $\pi$-criticality.} For $J_i, h_i$ near $0$ ($n_\pi \ll 1$), the IRFP distribution is similar to the static case, and the critical point can be understood in terms of domain walls (DWs) between regions where $J_i\gg h_i$ and those where $J_i\ll h_i$. Standard results show that in the fermionic language each DW binds a Majorana  state ${\tilde \gamma}^0_i$ at zero quasienergy, and the transition can be understood in terms of these. For $n_\pi\sim 1$, we again have a single IRFP distribution, but now centered at $\pi$. However, following~\cite{khemani_prl_2016} we may factor a global $\pi$ pulse from both terms of the drive, to recover the previous DW structure. Although still at zero quasienergy, here the DW Majoranas drive a transition between $\pi$SG-$0\pi$PM, owing to the global $\pi$-pulse. Again, the relevant parameter tuning the transition is the asymmetry between the distributions of $J_i$ and $h_i$ so the physics is essentially the same. 

Quite different physics arises for $n_\pi \sim 1/2$ where the couplings exhibit strong quenched spatial fluctuations between $0$ and $\pi$. 
This follows from the fact that there are {\it distinct} IRFP distributions for couplings near $0$ and $\pi$, such that the relevant critical physics is captured by a new class of ``$0\pi$-DWs'' unique to the Floquet setting. If $J_i$ is small and $h_i \sim \pi$ (consistent with $n_\pi \sim 1/2$), these correspond to DWs between $\pi$SG and PM regions, whereas if $h_i$ is small and $J_i \sim \pi$, the critical behavior can be understood in terms of DWs between SG and $0\pi$PM. In the fermion language
 each such $0\pi$ DW traps a Majorana bound state ${\tilde \gamma}^\pi_i$ at quasienergy $\pi$. This may also be deduced by comparing the edge modes of the adjacent phases (Fig.~\ref{fig:cartoons}a).  Since they are topological edge modes, a given $\pi$-Majorana trapped at a $0\pi$ DW can only couple to other $\pi$-Majoranas bound to $0\pi$ DWs, leading to a second emergent Majorana fermion chain whose dynamics are 
 independent from the initial chain (Fig.~\ref{fig:cartoons}b). If the intervening {puddles} are MBL, the tunneling between $\pi$-Majoranas is exponentially suppressed as $\sim {\rm e}^{-\ell}$, with the size $\ell$ of the puddles. Even if we start from a configuration where $J$ and $h$ are drawn from the same distribution, {there are still $\pi$-Majoranas bound to DWs separating infinite-randomness quantum critical regions where the couplings are near $0$ or $\pi$~\cite{supmat}, and} the typical tunnel 
  coupling is stretched exponential $\sim {\rm e}^{- \sqrt{\ell}}$~\cite{PhysRevB.51.6411,PhysRevLett.69.534}. Thus, the tunneling terms between the $\pi$-Majoranas remain short-ranged. Crucially, the criticality of this emergent $\pi$-Majorana chain is tuned by  $n_\pi$ (with $n_\pi=\frac{1}{2}$ at criticality), independently of the field-bond asymmetry that tunes the usual Ising transition. {We emphasize that although the universality class of this transition is still random Ising, it is described by flow towards an IRFP at $\pi$ quasienergy, and hence the spectral properties of the transition are distinct.} 

Observe that the  PM-$\pi$SG transition 
 involves the onset of TTSB, since the $\pi$SG is the prototypical example of a 
 time crystal. Similarly, the SG-$0\pi$PM transition involves the onset of TTSB at the ends of an open system. Therefore, we identify the $0\pi$ DWs as the degrees of freedom that are responsible for changes in TTSB.

\paragraph{RG treatment.} The above infinite randomness hypothesis suggests that the critical behavior at the dynamical Floquet transitions can be understood in terms of two  effectively static Majorana chains, one near quasienergy $0$ (${\tilde \gamma}^0_i$) and the other near quasienergy $\pi$ (${\tilde \gamma}^\pi_i$). While the criticality of the $0$ chain is driven by the asymmetry between $J$ and $h$ as in the usual Ising chain, the $\pi$ chain is critical for $n_\pi=\frac{1}{2}$ where there is a symmetry between $0$ and $\pi$ {couplings}. This picture can be confirmed explicitly~\cite{supmat} by considering instead the criticality of $F^2$, which should have couplings only near 0 and is described by an effectively static Hamiltonian $F^2 = {\rm e}^{-i 2 H_F}$. The dynamical properties of these two Majorana chains can be analyzed using standard RG techniques designed for static MBL Hamiltonians~\cite{PhysRevX.4.011052,PhysRevLett.112.217204,PhysRevLett.114.217201}. We decimate stronger couplings before weaker ones, putting the pair of Majoranas involved in the strongest coupling in a local eigenstate. Iterating this process leads to an IRFP which self-consistently justifies the strong disorder perturbative treatment. The resulting RG equations match those for the static random Ising model, except crucially, we can now have renormalization towards 0 or towards $\pi$ quasienergies in $F$ reflecting the decoupling of the two effective Majorana chains. This effective decoupling also persists in the presence of interactions ($U\neq 0$). Interactions within the $0$ and $\pi$ Ising chains flow towards 0 under RG much faster than the other couplings and are therefore irrelevant~\cite{PhysRevB.50.3799,PhysRevLett.112.217204}. While interactions also 
permit  Floquet-umklapp terms $\tilde \gamma^0_i \tilde \gamma^0_j \tilde \gamma^\pi_k \tilde \gamma^\pi_l$ that would couple the critical $0$ and $\pi$ chains at the multicritical point, such terms are also irrelevant, and so can be ignored as long as interactions are relatively weak~\cite{PhysRevB.45.2167,PhysRevB.50.3799,1402-4896-2015-T165-014040,PhysRevB.94.014205}. While weak interactions are irrelevant at the multicritical point, and very strong interactions are likely to drive thermalization, we leave open the possibility that intermediate interactions might drive the system to a new infinite-randomness critical point in the universality class of the random Ashkin-Teller model~\cite{1402-4896-2015-T165-014040}.

Therefore, for sufficiently weak interactions, the critical lines are always in the random Ising universality class. The four-phase multicritical point --- at which all four distributions are symmetric --- is in the Ising $\times$ Ising universality class.
This  picture of Floquet (multi) criticality  extends both symmetry-based reasoning used when all $h_i$ couplings are near $\pi$~\cite{PhysRevLett.118.030401}, and the analysis of the essentially static  $J_i,h_i \ll 1$~\cite{1742-5468-2017-7-073301} case.

\paragraph{Floquet (multi)criticality.} 
 
Combining this reasoning with standard IRFP results, 
we conclude that all the transitions show infinite-randomness Ising scaling:  
 the correlation length diverges as $\xi\sim |\Delta|^{-\nu}$ 
  where $\Delta$ characterizes the deviation from the critical lines, and $\nu=2$ or $1$ 
for average or typical quantities, respectively~\cite{PhysRevB.51.6411,PhysRevLett.69.534}.
This scaling should have universal signatures in dynamical (or eigenstate) correlation functions~\cite{PhysRevLett.69.534,PhysRevB.51.6411,PhysRevLett.112.217204,PhysRevLett.118.030401}, and in particular in the eigenstate entanglement entropy~\cite{PhysRevLett.93.260602,PhysRevB.90.220202,PhysRevLett.114.217201}. Assuming a system of size $L$ and open boundary conditions, the half-system entanglement entropy should scale with system size 
as $S_L \sim (\tilde c/6) \ln L$, up to nonuniversal additive contributions, with ``effective central charge'' $\tilde c=\ln 2/2$~\cite{PhysRevLett.93.260602}. At the multicritical point, we predict  $\tilde c = \ln2$ due to the criticality of the $0$ and $\pi$ Majorana chains. Our picture also predicts an emergent $\Z_2 \times \Z_2$ symmetry at the multicritical point, where the additional $\Z_2$ symmetry can be constructed explicitly as $D = F \sqrt{F^2}^\dagger$~\cite{pi-spin-glass,PhysRevLett.118.030401,PhysRevX.7.011026}. For a multicritical configuration with couplings near 0 or $\pi$, we find that $D$ is distinct from the original Ising symmetry of the model, and coincides with the fermion parity of the emergent $\pi$-Majorana chain,
\begin{equation}
\label{eq:majorana_D}
D = \prod_{j \in \{\rm 0\pi \ {\rm DWs} \}} \tilde \gamma^\pi_j.
\end{equation}
}

\paragraph{Numerics.}  

\begin{figure}
\includegraphics[width = \columnwidth]{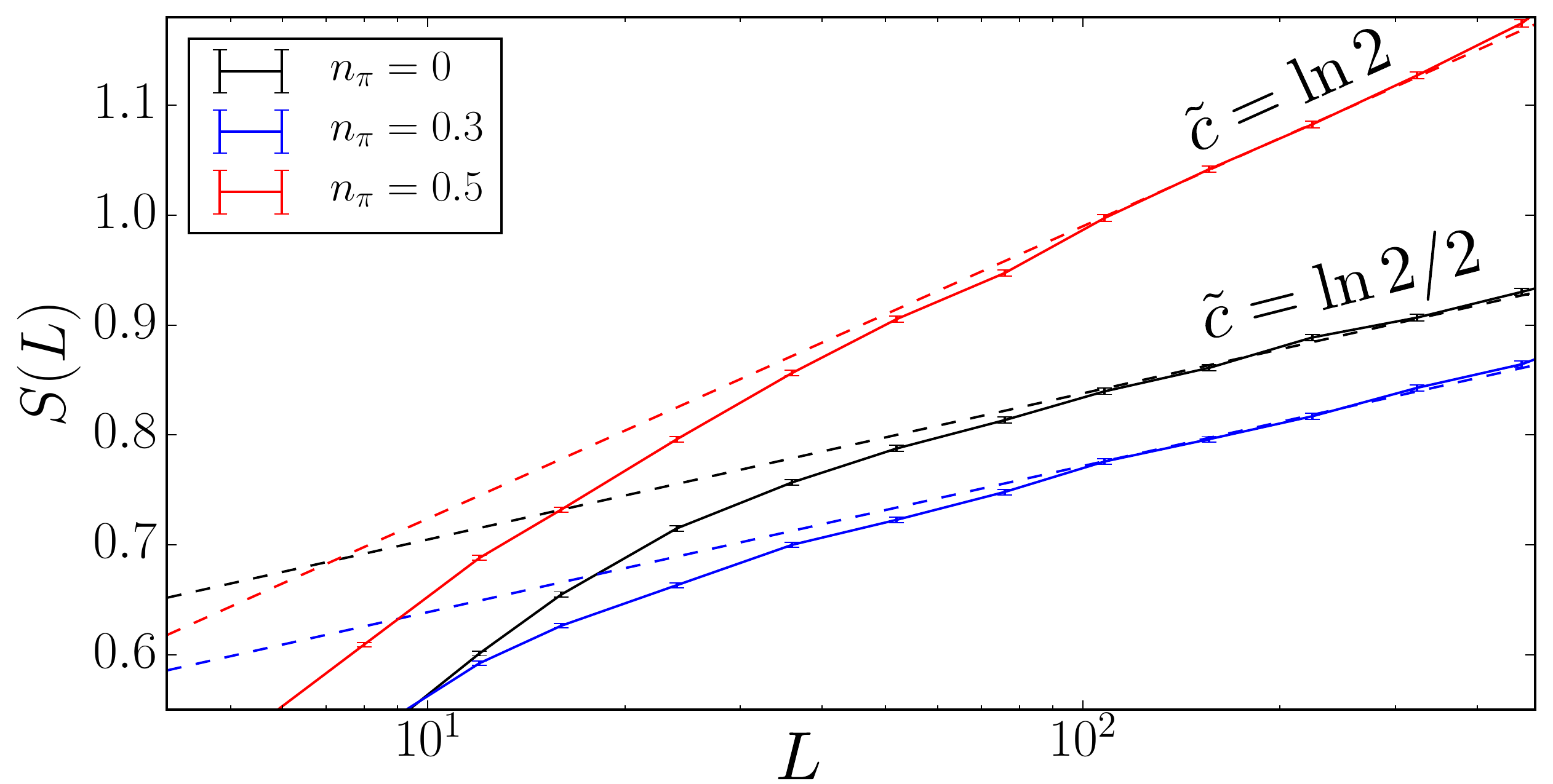}
\caption{\label{fig:entropy}{Scaling with system size of disorder- and eigenstate-averaged entanglement entropy $S$ for a cut at $L/2$. Dashed lines show predicted slopes for strong-disorder Ising criticality along the transition lines (blue, black); this doubles at the multicritical point (red).} }
\end{figure}

As stressed above, our picture of these transitions relies on the infinite randomness assumption. To justify this and to confirm our analytical predictions, we have performed extensive numerical simulations on the non-interacting model, leveraging its  free-fermion representation to access the full single-particle spectrum and to calculate the entanglement entropy of arbitrary eigenstates~\cite{supmat}. {
We average over 20,000 disorder realizations (with open boundary conditions), randomly choosing a Floquet eigenstate in each}.

 {Given our parametrization of disorder, the combination $\frac{1}{2}(n_\pi^h - n_\pi^J)$  provides a measure of the {asymmetry between $J$ and $h$ couplings}, while $\frac{1}{2}(n_\pi^h + n_\pi^J)$ measures the average probability of a $\pi$ coupling. Therefore, from our reasoning above and using} 
  the usual Ising duality, we expect a critical line for $n^\pi_J=n^\pi_h$. 
  {Combining} the Ising duality with Floquet symmetries leads to another critical line $n^\pi_J+n^\pi_h=1$ where we expect $0\pi$ infinite randomness behavior. Note that the bare disorder distributions are far from the infinite randomness fixed point expected to emerge at criticality. Nonetheless, as shown in Figure~\ref{fig:entropy}, we observe clear logarithmic scaling of entanglement along the self-dual lines $n^\pi_J=n^\pi_h$ and $n^\pi_J+n^\pi_h=1$ of Eq.~\eqref{eqModel}. We find $\tilde c \approx \ln 2 / 2$, consistent with the prediction that the lines are in the random Ising universality class. Deep in the phases, we find $\tilde c \approx 0$ consistent with the area-law scaling expected for Floquet MBL phases~\cite{1742-5468-2013-09-P09005,PhysRevLett.111.127201}. At the multicritical point $n_\pi^h = n_\pi^J = 1/2$, we find $\tilde c \approx \ln 2$, consistent with our expectation of two decoupled critical Ising chains. Although stability to quartic interchain couplings cannot be addressed in this noninteracting limit, we expect it on general grounds~\cite{PhysRevB.45.2167,PhysRevB.50.3799,1402-4896-2015-T165-014040}, modulo usual caveats on thermalization. Fig.~\ref{fig:phase_diagram}, showing the entanglement scaling across the entire phase diagram, summarizes these results. Finally, we have also numerically calculated the relative number of single particle quasienergies near 0 and near $\pi$, finding good agreement with a simple prediction from the infinite-randomness domain wall picture~\cite{supmat}. Moreover, Fig.~\ref{fig:phase_diagram} clearly shows that changing $n^\pi_J\pm n^\pi_h$ tunes across the critical lines, confirming that these parameters control  distribution asymmetries as in the IRFP picture (Fig.~\ref{fig:phase_diagram}, insets).

\paragraph{Experimental consequences.}
Let us now turn to some experimental consequences of the above predictions. Recent advances in the control of ultracold atomic arrays have brought models such as Eq.~\ref{eqModel} into the realm of experimental realizability~\cite{Kim:2010aa,Blatt:2012aa,Smith:2016aa}. The model hosts a time-crystal phase (the $\pi$ spin glass), the phenomenology of which has recently been directly observed~\cite{Choi:2017aa,Zhang:2017aa}. Even though, as mentioned earlier, these critical lines may eventually thermalize due to long-range resonances~\cite{PhysRevB.95.155129,PhysRevLett.119.150602,Ponte20160428}, the dynamics of the Ising universality class should persist through a prethermalization regime relevant to all reasonably accessible experimental timescales~\cite{PhysRevLett.115.256803,KUWAHARA201696}. Thus, the dynamical signatures of the transitions we have identified should be readily experimentally observable. 

One prominent experimental signature of this physics is the scaling behavior of the dynamical spin-spin autocorrelation function in Fourier space $C(\omega,t) \equiv \int_0^\infty d\tau e^{-i \omega \tau} \overline{\expec{\sigma_i^z (t+\tau) \sigma_i^z (\tau)}}$, with the overline representing a disorder average~\cite{PhysRevLett.118.030401}. In accordance with the random Ising universality class, the spin-spin autocorrelation function will scale as $\disavg{\expec{\sigma_i^z(t) \sigma_{i}^z(0)} } \sim {1}/{\log^{2-\phi} t}$~\cite{PhysRevLett.112.217204}, with the overline representing a disorder average and $\phi = (1 + \sqrt{5} ) / 2$ the golden ratio. Performing the Fourier transform, our analysis then predicts that along the $n^\pi_h = n^\pi_J$ critical line of the model, the Fourier peak at 0 quasienergy will decay as $C(0,t) \sim 1/\log^{2-\phi} t$; along the $n^\pi_h = 1 - n^\pi_J$ critical line the peak at $\pi$ quasienergy will decay the same way as $C(\omega/2,t) \sim 1/\log^{2-\phi} t$; and at the multicritical point, both peaks will decay in this way simultaneously, giving
\begin{equation}
C(0,t) \sim C(\omega/2,t) \sim \frac{1}{\log^{2-\phi} t}.
\end{equation} 
This slow, logarithmic decay, independently for the decoupled chains at $0$ and $\omega/2$, serves as an unambiguous signature of the universal multicritical physics we describe. The fact that the two decays are independent is highly nontrivial, since generic $\Z_2\times \Z_2$ multicritical points would have distinct scaling from either $\Z_2$ individually.

\paragraph{Discussion.} 
We have presented a generic picture of the transitions between MBL Floquet phases, and applied it to study the criticality of the periodically driven interacting random Ising chain. Our work can be generalized to more intricate Floquet systems, under the (reasonable) assumption that they flow to infinite randomness under coarse-graning. The resulting IRFP is enriched in the  Floquet setting: each distinct invariant Floquet quasienergy hosts an independent set of fixed-point coupling distributions. (For instance the $\Z_n$ model has $n$ such invariant quasienergies, $2\pi k/n$, with $k=1,\ldots,n$.)  
Systems at conventional IRFPs are tuned across criticality by adjusting the imbalance between distributions of  distinct couplings at the {\it same} quasienergy. At Floquet IRFPs, we may hold such single-quasienergy imbalances fixed and instead tune the imbalance between the distributions of couplings at {\it distinct} quasienergies. Transitions driven by such cross-quasienergy imbalances will usually involve an onset or change of TTSB in the bulk or at the boundary, and in this sense describe ``time crystallization''.  In some cases, it may be possible to leverage a Jordan-Wigner mapping in conjunction with these infinite-randomness arguments to arrive at a domain-wall description of the critical/multicritical physics. We anticipate that a variety of Floquet symmetry-breaking/symmetry-protected {topological} phases will be amenable to similar analysis, but we defer an exhaustive study to future work.


\paragraph{Materials and methods.} Numerical simulations were performed on the transverse-field Ising (TFI) chain, where we extract the entanglement entropy across a cut of length l from the boundary in an arbitrary eigenstate. We utilize the fact that the non-interacting TFI chain can be efficiently described as a system of free Majorana fermions~\cite{peschel_calculation_2003,eisler_evolution_2007}, details of which follow.

First, let us apply a Jordan-Wigner transformation to the TFI chain $\sigma_j^x = i \gamma_{2j} \gamma_{2j+1}$, $\sigma_j^y = \left( \prod_{l<j} i \gamma_{2l} \gamma_{2l+1} \right) \gamma_{2j+1}$, $\sigma_j^z = \left( \prod_{l<j} i \gamma_{2l} \gamma_{2l+1} \right) \gamma_{2j}$, where the $\gamma$ operators obey the Majorana algebra $\{ \gamma_{i},\gamma_{j} \} = 2\delta_{ij}$, $\gamma_{i}^2 = 1$, $\gamma_i^\dagger = \gamma_i$. This implies that $\sigma_j^z \sigma_{j+1}^z =i \gamma_{2j+1} \gamma_{2j+2}$. In the Majorana language, our periodically driven TFI Hamiltonian is 
\begin{equation}\label{eq:H}
H(t) = \begin{cases} 
H_1 = i \sum_{j=0}^{L-1} h_j \gamma_{2j} \gamma_{2j+1}  & 0 \leq t \leq T_1 \\
H_2 = i \sum_{j=0}^{L-2} J_j \gamma_{2j+1} \gamma_{2j+2}  & T_1 \leq t \leq T_1+T_2=T
\end{cases}
\end{equation}
where we set $T_1 = T_2 = 1/2$ for convenience. Now, if we are in a state satisfying Wick's theorem, the density matrix and all derived quantities are determined by the two-point correlator $C_{ij} = \expec{\gamma_i \gamma_j}$. The Majorana anti-commutation relation implies that $C_{ij} = 2 \delta_{ij} - C_{ji}$, so $C_{ij} = \delta_{ij} + a_{ij}$, where $a$ is some antisymmetric matrix. 

Let us first construct the Floquet evolution operator. To see how a Hamiltonian evolves the correlation function, first note that in the Heisenberg picture $C_{ij}(t) = \expec{\gamma_i(t) \gamma_j(t)}$. For $A$ and $B$ (distinct) Majorana operators, we have $e^{\alpha A B} = \cos \alpha + AB \sin \alpha$. Thus, $e^{\alpha AB} A e^{-\alpha AB} = A \cos 2 \alpha - B \sin 2 \alpha$, and $e^{\alpha AB} B e^{-\alpha AB} = B \cos 2 \alpha + A \sin 2\alpha$. If our Hamiltonian were not already in block-diagonal form, we would first need to pseudo-diagonalize it (in this case, perform a Schur decomposition) into the form $Q^T H Q = \sum_i \epsilon_i \gamma'_{2i} \gamma'_{2i+1}$, where $\gamma_i' = Q_{ij} \gamma_j$. Then, defining the block-diagonal matrix 
\begin{equation}
D(t) = \diag\left[\begin{pmatrix}
\cos (2\epsilon_k t) & -\sin (2\epsilon_k t) \\
\sin (2\epsilon_k t) & \cos (2\epsilon_k t)
\end{pmatrix}\right]_{k=1}^{2L}
\end{equation}
we find that $\gamma_i' (t) = D_{ij}(t) \gamma_j'(0)$. Defining $\Gamma (t) \equiv Q^T D(t) Q $, we see that the correlation function evolves simply as $C(t) = \Gamma(t) C(0) \Gamma(t)^T$. To construct the Floquet evolution operator, then, we write

\begin{equation}
F = U(T) = e^{-i T_2 H_2} e^{-i T_1 H_1} = Q_2^T D_2(T_2) Q_2 Q_1^T D_1(T_1) Q_1.
\end{equation}

For the simple drive considered above, $H_1$ and $H_2$ are already block-diagonal, so their $Q$ matrices are trivial and the exponentiation to construct each time evolution can be done explicitly. However, we do now need to perform a Schur decomposition on $F$ to bring it into block-diagonal form, given by a real orthogonal matrix $Q_F$. This rotates to the basis of single-particle Floquet eigenstates, with $\lambda_F^i$ the single-particle quasi-energies. 

The initial correlation function is simple in this basis: it will also be block diagonal with blocks of $\pm \sigma_x$, where the positive sign occupies the mode and the negative sign leaves it empty. Therefore, for an arbitrary Floquet eigenstate in the diagonal basis, $C'=\diag(\pm \sigma_x)_{i=1}^L$. We can then rotate back to the original variables by $C = Q_F C' Q_F^T$. 

Note that the $0$ and $\pi$ modes that emerge from our RSRG picture show up numerically as nearly degenerate states whose quasienergies must be resolved. In practice, this required implementing high-precision numerics -- beyond conventional machine precision -- which is described in more detail in the appendix~\cite{supmat}.

The above steps allow us to access the correlation function in any Floquet eigenstate. Let us now show how to calculate the entanglement entropy from that $C$ matrix. First, diagonalize the antisymmetric part of the correlation function $a = q^T \sigma q$ , where $q$ is orthogonal and $\sigma$ has form  $\sigma = \diag\begin{pmatrix}
0 & \lambda_i \\ -\lambda_i & 0 
\end{pmatrix}_{i=1}^L$. This can be achieved by a Schur decomposition, where the pseudo-eigenvalues are arranged such that $\sigma_{i,i+1} = \lambda_i$, and $\sigma^T = - \sigma$. Now define $\gamma' = q \gamma$. Then 
\begin{equation}
\expec{ \gamma'_{2k'-1} \gamma'_{2k }} = q_{2k'-1,i}q_{2k,j} (\delta_{ij} + \lambda_{k''} (q_{2k''-1,i}q_{2k'',j} - q_{2k'',i}q_{2k'',j}) ).
\end{equation}
From the orthogonality of $q$, $q_{\alpha i}q_{\beta i} = \delta_{\alpha\beta}$, so the only non-vanishing term is $q_{2k'-1,i} q_{2k,j} \lambda_{k''} q_{2k''-1,i} q_{2k'',j} = \lambda_k \delta_{kk'}\delta_{k'k''}$. Thus the only non-vanishing two-point function is
\begin{equation}
\expec{\gamma_{2k-1}' \gamma_{2k}'} = - \expec{\gamma_{2k}' \gamma_{2k-1}'} = \lambda_k.
\end{equation}
We can write this correlation function as arising from a single particle density matrix $\rho = \frac{1}{Z}\prod_k e^{ i \epsilon_k \gamma_{2k-1}' \gamma_{2k}' }$. 
Now, we can construct complex fermion operators from Majorana operators via $c_k = \frac{\gamma_{2k-1}' + i \gamma_{2k}'}{2}$, $c_k^\dagger = \frac{\gamma_{2k-1}' - i \gamma_{2k}'}{2}$, so $\gamma_{2k-1}' \gamma_{2k}' = -i(2 c_k^\dagger c_k - 1)$. This gives the density matrix as
\[
\rho = \prod_k \frac{e^{\epsilon_k (2 c_k^\dagger c_k - 1)}}{e^{\epsilon_k} + e^{-\epsilon_k}}.
\]
Thus the two-point function is $\expec{\gamma_{2k-1}'\gamma_{2k}'} = -i \expec{2 c_k^\dagger c_k - 1} = \lambda_k = -i \frac{e^{\epsilon_k} - e^{-\epsilon_k}}{e^{\epsilon_k} + e^{-\epsilon_k}} = -i \tanh \epsilon_k$. Now define $\mu_k = \abs{\lambda_k}$, giving $\epsilon_k = \tanh^{-1}(\mu_k)$. To find the entanglement entropy, write the density matrix as
\begin{equation}
\rho = \prod_k  \left[ p_k \proj{0_k} + (1-p_k) \proj{1_k} \right], \ \ \ \ p_k = \frac{e^{-\epsilon_k}}{e^{\epsilon_k} + e^{-\epsilon_k}}.
\end{equation}
Then the entanglement entropy in an arbitrary Floquet eigenstate is $S =- \Tr \rho \log \rho = -\sum_{k=1}^{L} p_k \log p_k + (1-p_k) \log (1-p_k)$.


\paragraph{Acknowledgments.} We thank Vedika Khemani, Shivaji Sondhi, Dominic Else, Chetan Nayak, Adam Nahum, Joel Moore, David Huse and especially Sthitadhi Roy for useful discussions, and are grateful to David Huse and Vedika Khemani for detailed comments on the manuscript. This work used the Extreme Science and Engineering Discovery Environment (XSEDE)\cite{john_towns_xsede:_2014}, which is supported by National Science Foundation grant number ACI-1053575. W.B. acknowledges support from the Department of Defense (DoD) through the National Defense Science \& Engineering Graduate Fellowship (NDSEG) Program, and from the Hellman Foundation through a Hellman Graduate Fellowship. We also acknowledge support from Laboratory Directed Research and Development (LDRD) funding from Lawrence Berkeley National Laboratory, provided by the Director, Office of Science, of the U.S. Department of Energy under Contract No. DEAC02-05CH11231, and the DoE Basic Energy Sciences (BES) TIMES initiative (M.K.); travel support from the California Institute for Quantum Emulation (CAIQUE) via PRCA award CA-15- 327861 and the California NanoSystems Institute at the University of California, Santa Barbara (W.B. and S.A.P.); support from NSF Grant DMR-1455366 at the University of California, Irvine (S.A.P.); and University of Massachusetts start-up funds (R.V.).

\bibliography{FloquetCriticality}


\bigskip

\onecolumngrid
\newpage



\section*{Supplementary material}

# Floquet Quantum Criticality: Supplemental Material

William Berdanier,[1, *] Michael Kolodrubetz,[1, 2, 3] S. A. Parameswaran,[4] and Romain Vasseur[5]

[1]*Department of Physics, University of California, Berkeley, CA 94720, USA*
[2]*Materials Sciences Division, Lawrence Berkeley National Laboratory, Berkeley, CA 94720, USA*
[3]*Department of Physics, The University of Texas at Dallas, Richardson, Texas 75080, USA*
[4]*The Rudolf Peierls Centre for Theoretical Physics, University of Oxford, Oxford OX1 3PU, UK*
[5]*Department of Physics, University of Massachusetts, Amherst, Massachusetts 01003, USA*
(Dated: September 21, 2018)

## I. NUMERICAL PRECISION AND CONVERGENCE

As argued in the main text, the disordered periodically driven Ising model flows to infinite randomness about both 0 and $\pi$. This leads to an abundance of single-particle modes exponentially near $E_F = 0$ and $E_F = \pi$, worsening with larger system sizes and stronger disorder. These near-degenerate modes give rise to a subtle numerical instability issue that requires going to high numerical precision, especially at strong disorder and/or large system sizes, in order to reliably calculate the correlation function and all derived quantities.

To that end, we implemented arbitrary precision[1] diagonalization in C++ using the Eigen[2] and MPFR[3] C++ libraries. We then checked convergence of our entropy calculations on a realization-by-realization basis by increasing the precision by hand. In Figure 1 we show a sample convergence plot for several disorder realizations at system size $L = 400$ Majoranas. Even at this relatively modest system size, precisions of $\epsilon < 10^{-20}$ are needed to reliably converge most disorder realizations. Data in the main text has been converged accordingly.

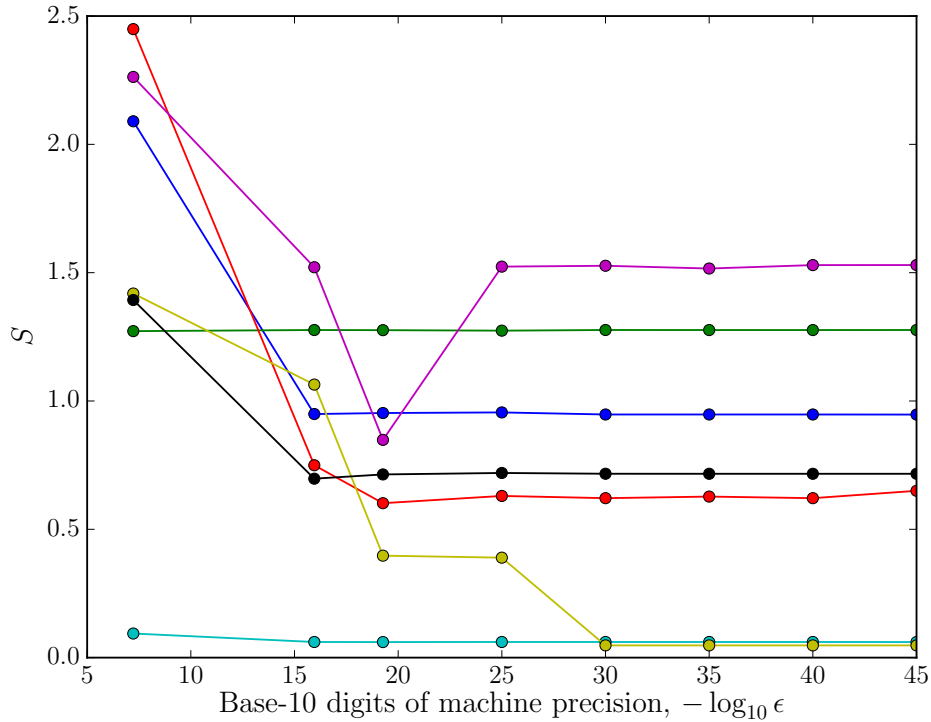

FIG. 1. Entanglement entropy converges with increased machine precision. Several disorder realizations (see main text) at system size $L = 400$ are plotted in various colors. Even for this relatively modest system size and box disorder, high precision is needed to accurately calculate the entanglement entropy. For comparison, float has $\epsilon \approx 10^{-8}$ and double has $\epsilon \approx 10^{-16}$.

## II. 0π ISING DUALITY, EMERGENT SYMMETRY, AND INFINITE RANDOMNESS STRUCTURE

In this appendix, we give a complete account of the 0π duality transformation described in the main text. We then calculate the associated emergent symmetry operator $D$ to second order in the weak coupling parameters $\epsilon, \delta \ll 1$, and show that it is simply the product of domain wall Majorana modes dressed by neighboring weak bonds, as argued in the main text. Finally, we detail a calculation of the ratio of the relative number of quasienergies near 0 to those near $\pi$ based on the infinite randomness domain wall structure from the main text and compare with numerical data.

### A. 0π-Ising duality and bond-field duality

First, consider the well-known bond-field duality of the static Ising model. This transformation defines dual spins on the bonds of the previous model, mapping $h_j \sigma_j^x \mapsto h_j \sigma_{j-1/2}^z \sigma_{j+1/2}^z$ and $J_j \sigma_j^z \sigma_{j+1}^z \mapsto J_j \sigma_{j+1/2}^x$. In the static case, this produces an Ising Hamiltonian with $h \leftrightarrow J$. Therefore this map does not preserve the spectrum, but does return a self-similar Hamiltonian. In the Floquet context, $F$ is actually not self-similar under this transformation. A bond-field transformation would send

$$F = e^{-\frac{i}{2}\sum_i J_i \sigma_i^z \sigma_{i+1}^z + U \sigma_i^z \sigma_{i+2}^z} e^{-\frac{i}{2}\sum_i h_i \sigma_i^x + U \sigma_i^x \sigma_{i+1}^x} \mapsto F' = e^{-\frac{i}{2}\sum_i J_i \sigma_i^x + U \sigma_i^x \sigma_{i+1}^x} e^{-\frac{i}{2}\sum_i h_i \sigma_i^z \sigma_{i+1}^z + U \sigma_i^z \sigma_{i+2}^z}, \quad (1)$$

where $F'$ is not in the same form as $F$ since the order of the field and bond terms is switched. However, $F'$ and $F$ are related by a unitary rotation: since $F$ is the product of two unitaries $F = AB$, conjugating either by $A$ or by $B^\dagger$ will give $F' = BA$. Also note that the transformation sending $h_i \mapsto -h_i$ without modifying $J$ is just a global $\pi$ rotation about the $y$- or $z$-axis, and hence does not affect the spectrum. Bond-field duality thus sends $h_i \mapsto \pm J_i$.

Unique to the Floquet case is the self-similar map $J_i \mapsto \pi + J_i$ and its bond-field dual $h_i \mapsto \pi + h_i$. For the first map, the Floquet operator transforms as

$$e^{-\frac{i}{2}\sum_i J_i \sigma_i^z \sigma_{i+1}^z + U \sigma_i^z \sigma_{i+2}^z} e^{-\frac{i}{2}\sum_i h_i \sigma_i^x + U \sigma_i^x \sigma_{i+1}^x} \mapsto (\prod_{\forall i} \sigma_i^z \sigma_{i+1}^z) e^{-\frac{i}{2}\sum_i J_i \sigma_i^z \sigma_{i+1}^z + U \sigma_i^z \sigma_{i+2}^z} e^{-\frac{i}{2}\sum_i h_i \sigma_i^x + U \sigma_i^x \sigma_{i+1}^x}, \quad (2)$$

using the identity $e^{i\frac{\pi}{2}\sigma_i^{(x,y,z)}} = i\sigma_i^{(x,y,z)}$, where we have dropped the overall phase factor because we are interested only in the operator content. Under periodic boundary conditions, $\prod_{\forall i} \sigma_i^z \sigma_{i+1}^z = 1$, so $F$ maps precisely onto itself. The dual transformation sends $F$ to $FG_{\text{Ising}}$, where $G_{\text{Ising}} = \prod_{\forall i} \sigma_i^x$ is the usual Ising symmetry operator. Since $[F, G_{\text{Ising}}] = 0$, this leaves the eigenstates invariant, but simply shifts their eigenvalues.

Composing the spectrum-preserving maps $(J_j, h_j) \mapsto (-J_j, -h_j) \mapsto (\pi - J_j, \pi - h_j)$ with bond-field duality gives the 0π-Ising duality transformation of $\sigma_j^z \sigma_{j+1}^z \mapsto (\pi - J_j) \sigma_{j+1/2}^x$, $h_j \sigma_j^x \mapsto (\pi - h_j) \sigma_{j-1/2}^z \sigma_{j+1/2}^z$ that maps the πSG to the PM and the SG to the 0πPM.

### B. Emergent symmetry and chain decoupling

Significant insight about this problem can be gleaned by considering not $F$ but rather $F^2$, in the same vein as Yao et al [4]. In particular, let us begin by considering a typical configuration of the model, where some couplings will be near 0 and others will be near $\pi$. As described in the main text, flow towards infinite randomness dictates that the multicritical point should be controlled by such chain configurations. Further, interaction terms are irrelevant, so in the limit of infinite randomness they have reached their fixed point of $U = 0$.

Let us work in the Majorana fermion picture, related to the spin picture via a Jordan-Wigner transformation, detailed in the Materials and Methods section. For convenience, let us relabel the two Majorana sublattices as $a_j = \gamma_{2j}$, $b_j = \gamma_{2j+1}$. In this language, the natural domains near the multicritical point are regions of Majorana couplings that are near $\pi$ ($J_i = \pi + \epsilon_i, h_i = \pi + \delta_i$) or near 0 ($J_j = \pi + \epsilon_j, h_j = \pi + \delta_j$), separated by domain walls. Considering a single domain wall, let all bonds to the left of $i = 0$ be near 0, and let all bonds to the right be near $\pi$. The Floquet operator in this case is then

$$F = e^{\sum_{i<0} \epsilon_i b_i a_{i+1} + \sum_{i \geq 0} (\pi/2 + \epsilon_i) b_i a_{i+1}} e^{\sum_{i<0} \delta_i a_i b_i + \sum_{i \geq 0} (\pi/2 + \delta_i) b_i a_{i+1}}, \quad (3)$$

where we have absorbed factors of $\frac{1}{2}$ into $\epsilon, \delta$. As a reminder, these Majorana operators obey the algebra $\{a_i, b_j\} = 2\delta_{ab}\delta_{ij}$, and $a^2 = b^2 = 1$. As can be straightforwardly shown, the Majorana fermion evolution operator is just $e^{\theta ab} = \cos\theta + ab\sin\theta$, where in particular, $e^{\frac{\pi}{2}ab} = ab$. We will drop overall phase factors to focus on operator content. Factoring out the $\pi$ pulses, then,

$$F = (\prod_{i\geq 0} b_i a_{i+1}) e^{\sum_i \epsilon_i b_i a_{i+1}} e^{\sum_i \delta_i a_i b_i} (\prod_{i\geq 0} a_i b_i)$$
$$= a_0 e^{-\epsilon_{-1} b_{-1} a_0 + \sum_{i\neq -1} \epsilon_i b_i a_{i+1}} e^{\sum_i \delta_i a_i b_i}, \tag{4}$$

where we have used the fact that if operators $A, B$ anticommute, then $e^B A = A e^{-B}$. If we now compute $F^2$, we get

$$F^2 = a_0 e^{-\epsilon_{-1} b_{-1} a_0 + \sum_{i\neq -1} \epsilon_i b_i a_{i+1}} e^{\sum_i \delta_i a_i b_i} a_0 e^{-\epsilon_{-1} b_{-1} a_0 + \sum_{i\neq -1} \epsilon_i b_i a_{i+1}} e^{\sum_i \delta_i a_i b_i}$$
$$= e^{\sum_i \epsilon_i b_i a_{i+1}} e^{\sum_{i\neq 0} \delta_i a_i b_i - \delta_0 a_0 b_0} e^{\sum_{i\neq -1} \epsilon_i b_i a_{i+1} - \epsilon_{-1} b_{-1} a_0} e^{\sum_i \delta_i a_i b_i}$$
$$= \exp\{2\sum_{i\neq -1} \epsilon_i b_i a_{i+1} + 2\sum_{i\neq 0} \delta_i a_i b_i + 2\sum_{\forall i} (\epsilon_{i-1}\delta_i b_{i-1} b_i - \epsilon_i \delta_i a_{i-1} a_i) - 4\epsilon_{-1}\delta_0 b_{-1} b_0 + \mathcal{O}(\lambda^3)\} \tag{5}$$

where $\epsilon_i, \delta_j$ are $\mathcal{O}(\lambda)$. To lowest order in BCH, both bonds touching $a_0$ cancel, *isolating this Majorana*. Now, as argued in [5], $F$ can be brought into the form $F = De^{-i\tilde{H}}$, where $\tilde{H}$ is a quasi-local Hamiltonian, and $D$ satisfies $D^2 = 1$ and $[D, \tilde{H}] = 0$ (hence $[D, F] = 0$). Now, $D$ both squares to 1 and commutes with $F$, as does $G_{\text{Ising}} = \prod_{\forall i} \sigma_i^x = \prod_{\forall j} ia_j b_j$. This lead us, as in the main text, to identify $D$ as a symmetry operator for $F$, with

$$D = F\sqrt{F^2}^\dagger. \tag{6}$$

This will only be well-defined if the branch cut of the square root operator at $-1$ is well-separated from the eigenvalues of $F^2$, which is indeed true in the limit of infinite randomness, since all eigenvalues of $F^2$ will be near 1. We can calculate $\tilde{H}$ and $D$ explicitly here using our formulas from above. The quasi-local Hamiltonian $\tilde{H}$ is

$$\tilde{H} = i\sum_{i\neq -1} \epsilon_i b_i a_{i+1} + i\sum_{i\neq 0} \delta_i a_i b_i + i\sum_{\forall i} (\epsilon_{i-1}\delta_i b_{i-1} b_i - \epsilon_i \delta_i a_{i-1} a_i) - 2i\epsilon_{-1}\delta_0 b_{-1} b_0 + \mathcal{O}(\lambda^3), \tag{7}$$

and the symmetry operator $D$ is

$$D = a_0 e^{-\epsilon_{-1} b_{-1} a_0 + \sum_{i\neq -1} \epsilon_i b_i a_{i+1}} e^{\sum_i \delta_i a_i b_i} e^{i\tilde{H}}$$
$$= a_0 e^{\delta_0 a_0 b_0 - \epsilon_{-1} b_{-1} a_0 + \epsilon_{-1}\delta_{-1} a_{-1} a_0 - \epsilon_0 \delta_0 a_0 a_1 + \mathcal{O}(\lambda^3)}$$
$$= e^{-\frac{1}{2}(\delta_0 a_0 b_0 - \epsilon_{-1} b_{-1} a_0 + \epsilon_{-1}\delta_{-1} a_{-1} a_0 - \epsilon_0 \delta_0 a_0 a_1 + ...)} a_0 e^{\frac{1}{2}(\delta_0 a_0 b_0 - \epsilon_{-1} b_{-1} a_0 + \epsilon_{-1}\delta_{-1} a_{-1} a_0 - \epsilon_0 \delta_0 a_0 a_1 + ...)}$$
$$= \tilde{a}_0, \tag{8}$$

where in the penultimate step we have used the fact that $a_0$ anticommutes with all terms in the exponential, and in the final step we have recognized that $\tilde{a}_0$ is simply $a_0$ dressed by a unitary rotation. Hence, $\tilde{a}_0$ is still a Majorana fermion operator, satisfying $\tilde{a}_0^2 = 1$ and $\tilde{a}_0^\dagger = \tilde{a}_0$. The exponential pieces on either side of $a_0$ are its exponential (or stretched exponential) tails. We identify $\tilde{a}_0$ as the emergent $\pi$ Majorana at this $0\pi$ domain wall.

For more generic domain wall configuration, we will see that a chain of Majoranas emerges whose dynamics are decoupled from the initial Majoranas. In particular, the emergent symmetry operator $D$ will take the form

$$D = \prod_{i \in \{\text{DWs}\}} \tilde{\gamma}_i \tag{9}$$

as claimed in the main text. Since domain walls must come in pairs in a chain with finite domain sizes, this will be just

$$D = \prod_j \tilde{\sigma}_j^x, \tag{10}$$

that is, the Ising symmetry operator (parity operator) of the chain formed by these domain wall Majoranas, up to an overall phase factor. These domain wall Majoranas are coupled by exponentially small tunneling terms: for a domain

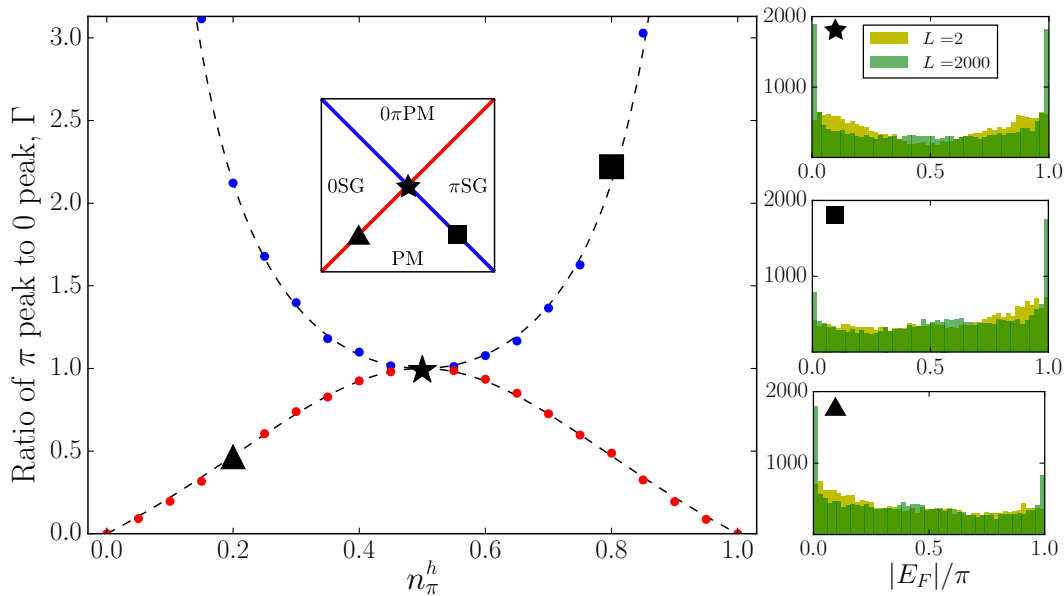

FIG. 2. Left: Relative number of single-particle quasienergies within 2% of $\pi$ to those within 2% of 0 ($\Gamma$ ratio), tracked numerically along the lines $n_\pi^J = n_\pi^h$ (red) and $n_\pi^J = 1 - n_\pi^h$ (blue), at system size $L = 2000$ and averaged over 100 realizations. The inset shows where these lines and the highlighted points (triangle, square, and star) lie on the phase diagram. The dashed line is the prediction from below. Right: Averaged quasienergy histograms showing the development of these peak heights around 0 and $\pi$ with increasing system size.

of size $L$, we must go to order $L$ in BCH to see a coupling term appear in the exponent. For $0/\pi$ regions deep in a phase, this coupling is thus exponentially small in $L$. For critical regions, exponential scaling is replaced by stretched exponential scaling, but the above is otherwise the same.

Finally, we note that our picture of two decoupled chains can be mapped explicitly onto the disordered $XY$ model, with interactions between the chains promoting this to an $XYZ$ model. Following a Jordan-Wigner transformation, we see that $X_i X_{i+1} = i b_i a_{i+1}$ and $Y_i Y_{i+1} = i a_i b_{i+1}$, where $a$ and $b$ are the two Majorana fermion sublattices. The even $J_{XX}$ couplings and odd $J_{YY}$ couplings then form one Majorana chain, and the remaining couplings form the other. If the $J_{XX}, J_{YY}$ couplings are drawn from the same distribution, these two Majorana chains are critical. The flow of interactions under renormalization has been studied in [6], where they found the $J_{ZZ}$ interaction terms to be irrelevant.

### C. Infinite randomness structure and domain wall quasienergies

Here we detail a simple calculation of the relative number of states with single particle quasienergies at 0 and $\pi$ based on our infinite randomness arguments in the main text, and compare with numerics. We utilize the parameters $n_\pi^{h,J}$ introduced in the main text. Define the ratio of the number of single particle quasienergies within some small number $\epsilon$ of $\pi$ to the number within $\epsilon$ of 0 to be $\Gamma_\epsilon$. First, let us move along the line $n_\pi^h = n_\pi^J$. In our infinite randomness picture, we claim that each domain wall should host a single-particle mode at quasienergy $\pi$, and the other quasienergies should be near 0. The probability of a domain wall is $n_{\rm DW} = 2 n_\pi^h (1 - n_\pi^h)$, so we predict that along this line, taking $\epsilon \to 0$, $\Gamma(n_\pi^h) = \frac{n_{\rm DW}}{1 - n_{\rm DW}} = \frac{1}{2 n_\pi^h (n_\pi^h - 1) + 1} - 1$. Moving along the other critical line, $n_\pi^h = 1 - n_\pi^J$, we expect each domain wall to host a mode at quasienergy 0, and hence the $\Gamma$ ratio should be $\Gamma(n_\pi^h) = \frac{1 - n_{\rm DW}}{n_{\rm DW}} = \frac{1}{2 n_\pi^h} + \frac{1}{2(1 - n_\pi^h)} - 1$. These predictions are compared with numerics in Figure 2, finding good agreement. Therefore, even though we start far from the infinite randomness fixed point, the energetics are nonetheless controlled by the infinite randomness domain wall structure.

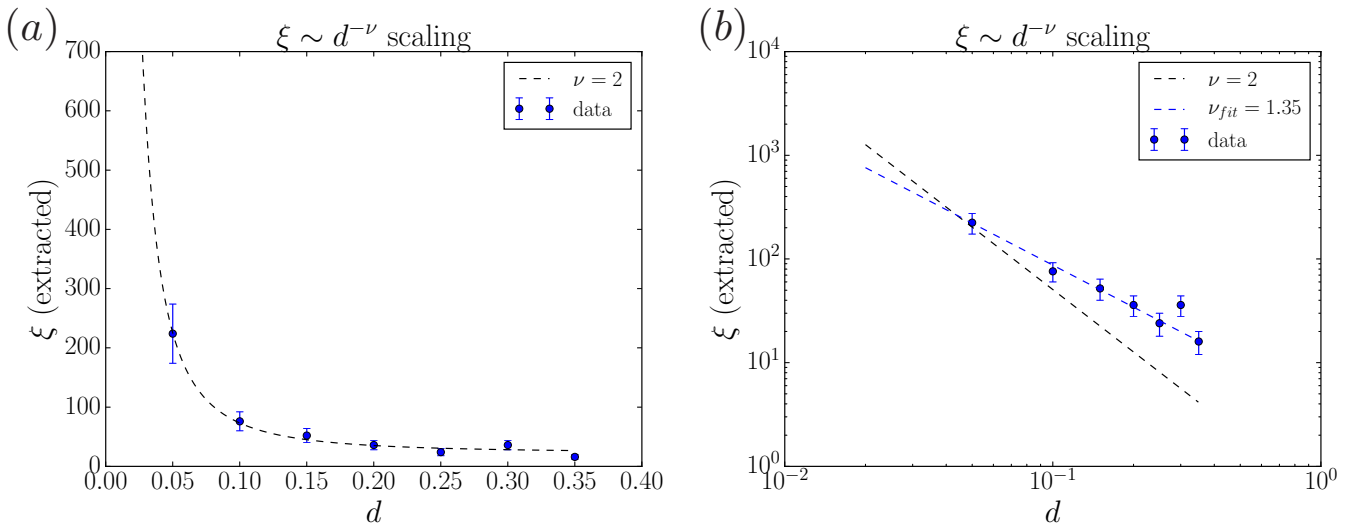

FIG. 3. Scaling of the correlation length with distance from criticality. We define the "correlation length" $\xi$ as the length at which the disorder-averaged entanglement entropy $\overline{S}(l)$ crosses over from logarithmic growth to a constant (area-law), i.e. the length when $\overline{S}(l)$ saturates to within $\varepsilon = 1.5\%$ (see text). We take the path $(n_h^\pi, n_J^\pi) = (0.4, 0.4 - d)$ here. (a) Scaling of $\xi$ with $d$ (blue dots). The black dashed line is a fit to the expected scaling form $\xi = a \times d^{-\nu} + b$, with $a$ and $b$ fit parameters and $\nu = 2$. (b) The same data on a log-log plot. One can see power-law scaling, with fit exponent $\nu_{\text{fit}} \approx 1.35$; given the complexities of the IRFP scaling and disorder averaging, we would expect $\nu_{\text{typ}} = 1 < \nu_{\text{fit}} < 2 = \nu_{\text{avg}}$ from a coarse measure like entanglement.

## III. MICROSCOPIC JUSTIFICATION FOR INFINITE RANDOMNESS

In this appendix, we provide further evidence for flow to an infinite randomness fixed point. First, we extract a "correlation length" from the scaling of the average entanglement, and see how this quantity scales with the distance from criticality. Second, we present results of an upcoming article that gives a microscopic justification of flow to infinite randomness based on an explicit strong-disorder renormalization group procedure [7].

One of the pillars of our argument in the main text is the assumption that critical points between Floquet MBL phases are controlled by infinite randomness fixed points (IRFPs) of a strong-disorder renormalization group. From this assumption, we derived several nontrivial predictions, one of which is the (open boundary condition) entanglement entropy scaling $S(L) \sim (\tilde{c}/6) \log L$, with $\tilde{c} = (1/2) \ln 2$ along the critical lines of the driven disordered Ising model $\tilde{c} = \ln 2$ at the multicritical point. Indeed, entanglement entropy scaling is one of the best and most reliable indicators of an infinite randomness fixed point, as it is self-averaging and allows for direct extraction of the disordered central charge, which identifies the universality class of the fixed point [8, 9]. We found good numerical agreement with these predictions in the main text.

Here we provide further numerical evidence of this IRFP by analyzing the scaling of the entanglement correlation length with distance from criticality. At an IRFP, we generically expect power-law scaling of a correlation length $\xi$ with distance from criticality $d$ as $\xi \sim d^{-\nu}$, where $\nu = 1$ for a typical quantity (i.e, $e^{\overline{\log \mathcal{O}}}$) and $\nu = 2$ for an average quantity (i.e. $\overline{\mathcal{O}}$) [10, 11]. Given that our analysis has focused on the entanglement entropy, and that spin-spin correlation functions are difficult to extract in the Majorana basis as they involve arbitrarily-long 'string' operators, we extract a correlation length directly from the entanglement data (see Figure 3). In particular, in a one-dimensional MBL phase, we expect the entanglement entropy to cross over from logarithmic growth to a constant, satisfying an entanglement 'area-law.' We can numerically estimate the length scale of this crossover as the length $\xi$ at which the disorder-averaged entanglement entropy saturates to within some numerical threshold $\epsilon$. In Figure 3, we set this threshold at $\epsilon = 1.5\%$, and calculate this crossover length $\xi$ along the line $(n_h^\pi, n_J^\pi) = (0.4, 0.4 - d)$. Since the entanglement is a disorder-averaged quantity, we would expect $\nu > 1$; given the immense length scales required to see the long-distance Griffiths effects that distinguish average from typical quantities, it does not numerically saturate to $\nu = 2$. We instead expect some intermediate value of $2 > \nu > 1$, as has been found in the literature; for instance, Ref. [4] found $\nu = 1.3$. We find $\nu_{\text{fit}} = 1.35$, in accordance with this expectation.

Next, we present some results of an upcoming article giving a microscopic justification of flow to infinite randomness [12]. In this work, we introduce a generalization of the Schrieffer-Wolff transformation to Floquet unitary

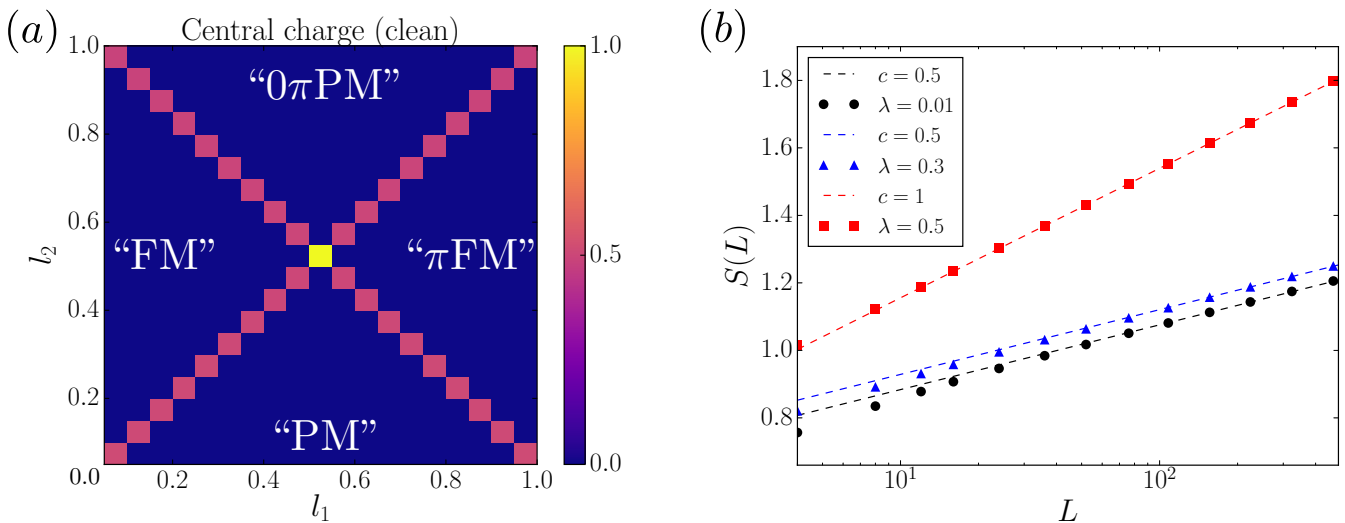

FIG. 4. Criticality in the clean (non-disordered) model. (a) Phase diagram of the clean model showing the central charge extracted from entanglement scaling $S \sim \frac{c}{3} \log L$ in the Floquet ground state (all negative quasienergies filled). Unlike the high temperature spin glass phases in the main text, these phases are more accurately labeled as ferromagnets or paramagnets, as they behave like ground states of the Ising model. We have used quotation marks to indicate that such phases are not robust to heating, unlike their counterparts in the disordered system. Along the critical lines of the diagram the Floquet ground state is described by a CFT with central charge $c = 1/2$, while the multicritical point shows $c = 1$. The phases saturate to area-law (constant) entanglement. (b) Entropy slices at the points $l_1 = l_2 = \lambda$ in the diagram to the left. $\lambda = 0$ and $\lambda = 0.3$ both show $c = 1/2$, while the multicritical point $\lambda = 0.5$ shows $c = 1$.

operators, allowing for decimation of 'strong' bonds in much the same way as in the static problem, with the strong bond treated non-perturbatively and the rest of the chain a perturbation. In particular, we find that for decimations within a domain of bonds all near 0 or near $\pi/2$ (within a 0 or $\pi$ domain), decimating a strong bond $\Omega i \gamma_1 \gamma_2$ with neighbors $J_L i \gamma_0 \gamma_1$ and $J_R i \gamma_2 \gamma_3$ gives a renormalized coupling $\tilde{J} i \gamma_0 \gamma_3$ with the rule

$$\tan \tilde{J} \approx \tilde{J} = \frac{J_L J_R}{\tan \Omega}. \tag{11}$$

Note that this reproduces the static rule of $\tilde{J} = J_L J_R / \Omega$ when $\Omega \ll 1$, as it must. Within a $\pi$ domain, the renormalized bond is $\pi/2 + \tilde{J}$. Rewriting this rule in terms of logarithmic variables, $\log \tilde{J} = \log J_L + \log J_R - \log \tan \Omega$, the distributions flow to an infinite randomness fixed point by analogy with the analysis from Fisher [10, 11]. We also find that care must be taken at the domain walls, as suggested by the picture in the main text that these domain walls host topological edge modes. We find that the decimations involving domain wall Majorana operators naturally lead to two decoupled Majorana chains, one near 0 and the other near $\pi$ quasienergy, that can independently be tuned to criticality – precisely the result from the main text. This analysis provides a microscopic justification for our more coarse-grained arguments, further justifying a flow to infinite randomness.

## IV. CLEAN (NON-DISORDERED) CRITICALITY

In this appendix, we discuss the criticality of the clean (non-disordered) driven Ising model. In the absence of interactions, this model still has a non-trivial phase structure [13, 14]. For clarity, the clean model is defined by the Floquet evolution operator

$$F = e^{-i\frac{\pi}{2} \sum_i l_2 \sigma_i^z \sigma_{i+1}^z} e^{-i\frac{\pi}{2} \sum_i l_1 \sigma_i^x}, \tag{12}$$

where $l_1, l_2$ are normalized variables defined modulo 1. In the clean model, we do not expect a flow to infinite randomness, and thus cannot apply the picture of criticality from the main text, which relied on viewing a typical configuration of the chain at criticality as composed of adjoining regions deep in the neighboring phases. Nonetheless, the clean case still displays critical behavior, described instead by a conformal field theory (CFT). This is possible

because the clean case, in the absence of interactions, is integrable; with interactions we expect that the model displays only prethermal phases and critical lines, and will eventually thermalize to infinite temperature [15].

We note that the critical lines of the clean model are set entirely by the dualities mentioned in the main text and explained in Appendix II. Bond-field duality, as in the static clean Ising model, implies that the self-dual line of $l_1 = l_2$ will be critical. In the driven case, $0\pi$ Ising duality maps $l_1 \sigma_i^x \mapsto (1 - l_1)\sigma_i^z \sigma_{i+1}^z$ and $l_2 \sigma_i^z \sigma_{i+1}^z \mapsto (1 - l_2)\sigma_i^x$. This changes the order of the pulses, but as discussed above, since $F$ is of the form $F = AB$ with $A, B$ unitary operators, $F' = BA$ is related to $F$ by a simple unitary transformation and hence the spectrum is unaffected. A second self-dual and hence critical line is then $l_1 = 1 - l_2$. These two self-dual lines cross at the multicritical point $(l_1, l_2) = (0.5, 0.5)$.

As mentioned in the main text, Floquet systems generally have no notion of a ground state, since quasi-energy is only defined modulo $2\pi/T$ and hence lives on a circle. Nonetheless, for weak enough driving, one state will be adiabatically connected to the ground state – usually deemed the "Floquet ground state" (see, e.g., [16]). In our model, this state corresponds to simply filling all single-particle modes of negative quasi-energy. In the weak-driving limit, all quasi-energies will be near 0, and hence the problem will be nearly static; the Floquet ground state is then just the ground state of the static model ($T \to 0$). We expect that in the $T \to 0$ limit that the Floquet ground state should be described by a CFT with central charge $c = 1/2$, since this limit recovers a static critical Ising model [17].

In Figure 4, we show numerical calculations of the entanglement entropy in the Floquet ground state from which we can extract the central charge $c$: with periodic boundary conditions, we expect the entanglement entropy to satisfy $S \sim \frac{c}{3} \log L$ [18]. We find that along the critical lines of the diagram, the Floquet ground state gives $c = 1/2$ while the multicritical point shows $c = 1$.

---



\end{document}